\documentstyle[12pt,fleqn]{article}
\begin{document}
\baselineskip=7.8mm

\begin{center}
{\large\bf  Energy Sources of Soft Gamma-Ray Repeaters}

\vspace{5.0mm}
K. S. Cheng$^1$ and Z. G. Dai$^2$ 

$^1${\em Department of Physics, University of Hong Kong, Hong Kong, 
China}

$^2${\em Department of Astronomy, Nanjing University, Nanjing 210093,
China}

\end{center}

\vspace{3mm}

\begin{center}
ABSTRACT
\end{center}

Quiescence and burst emission and relativistic particle
winds of soft gamma-ray  repeaters (SGRs) 
have been widely interpreted to result from
ultrastrongly magnetized neutron stars. In this magnetar model,
the magnetic energy and gravitational energy of the neutron stars 
are suggested as the energy sources of all the emission and winds. 
However, Harding, Contopoulos \& Kazanas (1999) have shown that the
magnetic field should be only $\sim 3\times 10^{13}$ G in order to 
match the characteristic spin-down timescale of SGR $1806-20$ 
and its SNR age. Here we argue that if the magnetic 
field is indeed so weak, the previously suggested energy sources seem
problematic. We further propose a plausible model in which SGR pulsars
are young strange stars with superconducting cores 
and with a poloidal magnetic
field of $\sim 3\times 10^{13}$ G. In this model, the movement of the
flux tubes not only leads to crustal cracking of the stars, giving rise 
to deconfinement of crustal matter to strange matter, but also to 
movement of internal magnetic toruses with the flux tubes. 
The former process will result in burst and quiescence emission and the
latter process will produce steady relativistic winds which will power 
the surrounding supernova remnants.

\noindent
{\em Subject headings:} gamma-ray: bursts -- magnetic fields 
-- dense matter -- stars: neutron 

\newpage

\begin{center}
1. INTRODUCTION
\end{center}

The soft gamma-ray repeaters (SGRs) are a small, enigmatic class of
high-energy transient sources, which differ from classical gamma-ray
bursts by their durations (typically $0.1-1$ s), soft spectra with 
characteristic energies of $\sim 30-50$ keV and their repetition.
The other properties of four known SGRs include: 
(1) all of them are associated with supernova remnants (SNRs). 
SGR $0525-66$ appears to be associated with SNR N49 in the Large
Magellanic Cloud (Evans et al. 1980; Cline et al. 1982). 
The second burster, 
SGR $1806-20$, which produced $\sim 110$ bursts during a 7-yr span 
(Laros et al. 1987) and recently became active again (Kouveliotou et al.
1994), appears to be coincident with SNR G$10.0-0.3$ (Murakami et al.
1994), confirming an earlier suggestion (Kulkarni \& Frail 1993). 
The age of this SNR was estimated to be $\sim 10^4$ yr based on angular
diameter versus surface brightness argument (Kulkarni et al. 1994).
The third burster, SGR $1900+14$, is associated with SNR G$42.8+0.6$
(Vasisht et al. 1994), whose age is also $\sim 10^4$ yr. The fourth
burster, SGR $1627-41$, was recently discovered to be associated with
SNR G$337.0-0.1$ (Hurley et al. 1999; Woods et al. 1999). From these
SGR-SNR associations, the burst peak luminosities can be estimated to
be a few orders of magnitude higher than the standard Eddington luminosity
for a stellar-mass star. For example, SGR $1806-20$ produced bursts with
$\sim 10^4$ times the Edditington luminosity (Fenimore, Laros \& Ulmer
1994). (2) In addition to short bursts of soft gamma-ray photons, the
persistent X-ray emission has been detected from SGRs (Murakumi et al. 
1994; Vasisht et al. 1994; Rothschild, Kulkarni \& Lingenfelter 1994).
The luminosities of the persistent X-ray emission are $\sim 7\times
10^{35}\,{\rm erg}\,{\rm s}^{-1}$ for SGR $0525-66$, $\sim 3\times 
10^{35}\,{\rm erg}\,{\rm s}^{-1}$ for SGR $1806-20$, and $\sim
10^{35}\,{\rm erg}\,{\rm s}^{-1}$ for SGR $1900+14$. (3) Recently,
a period of $P=7.47$ s and its derivative $\dot{P}=8.3\times 
10^{-11}\,{\rm s}\,{\rm s}^{-1}$ have been detected from SGR $1806-20$
in quiescent emission (Kouveliotou et al 1998), and a period of $P=5.16$ s
and its derivative $\dot{P}=6.0\times 10^{-11}\,{\rm s}\,{\rm s}^{-1}$ 
have been discovered from SGR $1900+14$ in quiescent emission
(Kouveliotou et al. 1999). All of these observations clearly show that
SGRs are young pulsars. Furthermore, if the period derivatives are
driven by magnetic dipole radiation, it can be shown (Pacini 1969) 
that the dipolar magnetic field is given by $B_p=3.2\times 10^{19}
(P\dot{P})^{1/2}$ G, which would yield dipolar magnetic fields
of $8\times 10^{14}$ and $5\times 10^{14}$ G for SGR $1806-20$ and
SGR $1900+14$ respectively. Therefore, the SGR pulsars are magnetars,
``neutron stars" with magnetic fields $\ge 10^{14}$ G. Such stars were
first proposed by Duncan \& Thompson (1992), Usov (1992) and 
Paczy\'nski (1992). 

However, the above estimate of dipolar magnetic fields leads to
characteristic spin-down ages much smaller than the SNR ages. This
difficulty can be alleviated by 
introducing relativistic particle outflows from SGRs.
The existence of such a wind has been inferred indirectly by X-ray and
radio observations of the synchrotron nebula G$10.0-0.3$ around
SGR $1806-20$ (Murakami et al. 1994; Kulkarni et al. 1994). 
Thompson \& Duncan (1996) estimated that the particle luminosity
from SGR $1806-20$ is of the order of $10^{37}\,{\rm erg}\,{\rm s}^{-1}$.
Such an energetic wind will also affect the spin-down torque
of the pulsar by distorting the dipole field structure near the light
cylinder (Thompson \& Blaes 1998). Furthermore, Harding, Contopoulos
\& Kazanas (1999) have found that if SGR $1806-20$ puts out a continuous 
particle wind of $10^{37}\,{\rm erg}\,{\rm s}^{-1}$, then the pulsar
age is consistent with that of the surrounding SNR, but the derived
surface dipole magnetic field is only $3\times 10^{13}$ G, in the 
range of normal radio pulsars.

It has been widely thought that ultra-strong magnetic fields are 
an origin of SGR quiescence and burst emission and relativistic particle
winds (Thompson \& Duncan 1995, 1996). As analyzed in Section 2,                
however, a magnetic field of $3\times 10^{13}$ G may be too weak
to be considered as an energy source of SGR quiescence emission and 
relativistic particle winds. Furthermore, the rotational energy,
gravitational energy and crustal strain energy of SGR pulsars are 
not yet suitable. Following Cheng \& Dai (1998) and Dai \& Lu (1998), 
in Section 3 we will propose a model in which SGR pulsars are young, 
magnetized strange stars with superconducting cores. We argue that
this model can provide an explanation for all the observed properties 
of SGRs including steady winds with luminosities of $\sim 10^{37}\,
{\rm erg}\,{\rm s}^{-1}$. In the final section, we will discuss some
differences between anomalous X-ray pulsars (AXPs) and SGRs in our model.

\begin{center}
2. PREVIOUSLY SUGGESTED ENERGY SOURCES
\end{center}

Observationally, SGRs have both quiescence and burst emission 
and steady winds. In the following we estimate the energies 
of the emission and wind from SGR $1806-20$. First, assuming that 
the luminosity of the wind is $L_w\sim 10^{37}\,{\rm erg}\,{\rm s}^{-1}$,
we obtain the total observed wind energy 
$E_w=L_wt_{\rm SNR}\sim 3\times 10^{48}\,{\rm ergs}$. 
Second, the total energy of the persistent X-ray emission 
is given by $E_x=L_xt_{\rm SNR}\sim 6\times 10^{46}\,{\rm ergs}$,
where $L_x$ is the persistent X-ray luminosity ($\sim 2\times 10^{35}\,
{\rm erg}\,{\rm s}^{-1}$). Third, the total observed energy of 
SGR bursts, assuming isotropic emission, can be estimated by
$E_{b,{\rm tot}}=E_b(t_{\rm SNR}/\tau_{\rm int})\sim 3\times 10^{46}
(E_b/10^{41}{\rm ergs})(t_{\rm int}/10^6{\rm s})^{-1}$,
where $E_b$ is the typical energy of a burst ($\sim 10^{41}\,
{\rm ergs}$) and $\tau_{\rm int}$ is the interval timescale of
SGR bursts ($\sim 10^6$ s).

Theoretically, there are four energy sources for the persistent X-ray
and burst emission and the wind. The first energy source is the rotational
energy of the pulsar $E_{\rm rot}\sim 4\times 10^{44}(P/7.47\,
{\rm s})^{-2}\,{\rm ergs}$. Second, assuming a uniform poloidal field
configuration in the interior, the total magnetic energy is 
$E_B\sim 3\times 10^{44}[B_p/(3\times 10^{13}{\rm G})]^2\,{\rm ergs}$.
Moreover, the numerical studies of Heyl \& Kulkarni (1998) show that
a magnetic field with $\sim 3 \times 10^{13}$ G doesn't obviously decay
even in $10^6$ yr, implying that this magnetic energy cannot be varied
in the SGR age. The third energy source is the gravitational energy 
of the pulsar. It is well known that the available gravitational 
energy for a rotating star ($\Delta E_G$) is only the difference 
in the gravitational energy between this star and a nonrotating 
(spherical) star for the same baryon mass. Assuming that the SGR 
pulsar is a slowly rotating Maclaurin spheroid, we easily demonstrate 
$\Delta E_G = 5E_{\rm rot}\sim 2\times 10^{45}(P/7.47\,{\rm s})^{-2}\,
{\rm ergs}$. The final energy source is the crustal strain energy. 
Assuming that the SGR pulsar is a neutron star, we obtain the strain 
energy (Baym \& Pines 1971): $E_{\rm strain}\sim 2\times 10^{45}
\epsilon^4\,{\rm ergs}\ll \Delta E_G$, 
where $\epsilon$ is the eccentricity of the star. 

Comparing the theoretical energy sources with the observed energies 
of the persistence and burst emission and the wind, we find that 
the previously suggested energies are much smaller than required
by observations. For example, the magnetic energy is about four 
orders of magnitude smaller than the wind energy. Therefore, 
we conclude that these energy sources are too weak to be considered 
as origins of SGRs. Furthermore, even if a magnetar-strength field 
of $\sim 10^{14}$ G, as Harding et al. (1999) argued in the case of 
an episodic wind with small duty cycle, is assumed, this conclusion 
remains correct. What are energy sources of SGRs?

\begin{center}
3. OUR ENERGY SOURCES
\end{center}
 
We now propose a plausible model for SGRs, in which SGR pulsars 
are young, magnetized strange stars with superconducting cores. 
The structure of strange stars has been widely studied (for a recent
review see Cheng, Dai \& Lu 1998). An interesting possible signature 
for the existence of strange stars has been found in a few low-mass 
X-ray binaries (Stergioulas, Kluzniak \& Bulik 1999), in which the kHz
quasi-periodic oscillation phenomena were recently observed (Zhang et al. 
1998). Another strange star candidate is an unusual hard X-ray
burster, GRB J1744$-$28 (Cheng et al. 1998).  

It is well known that supernova explosions are very likely to 
produce neutron stars. Because of hypercritical accretion, the neutron
stars may subsequently accrete sufficient mass ($\sim 0.5M_\odot$) 
to convert to massive strange stars (Cheng \& Dai 1996; 
Dai \& Lu 1999; Wang et al. 1999). 
Since the density profile of a strange star is much
different from that of a neutron star for the same baryon mass,
differential rotation may occur in the interior of the newborn strange
star. Dai \& Lu (1998) have argued that such a differentially
rotating strange star could lead to a classical gamma-ray burst.
The basic idea of their argument is: In a differentially rotating 
strange star, internal poloidal magnetic field will be 
wound up into a toroidal configuration and linearly amplified as one 
part of the star rotates about the other part. Only when it increases 
up to a critical field, $B_f\sim 2\times 10^{17}$ G, 
will the toroidal field be sufficiently buoyant to overcome 
fully the stratification in the composition of the strange 
star core. And then the buoyant magnetic torus will be able to float 
up to and break through the stellar surface. Reconnection of the surface 
magnetic field will produce a quickly explosive event as a peak of
a gamma-ray burst. This idea is similar to that of Kluzniak \& Ruderman 
(1998) who discussed the neutron star case. Here we further suggest 
that after the gamma-ray burst many magnetic toruses with $B_\phi <B_f$ 
(toroidal field configuration) could remain in the interior of 
the strange star in a timescale of $\sim 10^4$ yr. 

After its birth, a strange star must start to cool due to neutrino
emission. As with a neutron star, the strange star core may become 
superconducting when its interior temperature is below the critical
temperature. Bailin \& Love (1984) found that the superconducting
transition temperature in strange matter is about 400 keV. Therefore,
a strange star with age of $\sim 10^4$ yr after its supernova 
birth must have a core temperature much lower than the superconducting
transition temperature. The interior temperature of the strange star 
decreases as $T\approx 10^8(t/{\rm yr})^{-1/4}$ K, so $T\sim 10^7$ K
when $t\sim 10^4$ yr. The quark superconductor is likely to be 
marginally type-II with zero temperature critical field 
$B_c\sim 10^{17}$ G (Bailin \& Love 1984; Benvenuto, Vucetich 
\& Horvath 1991; Chau 1997). Furthermore, Chau (1997) argued that after 
the quark superconductor appears in the strange star, the coupling 
between quantized vortex lines and (poloidal) magnetic flux tubes
in the strange star is so strong that when the vortex lines are moving
outward due to spinning down of the star, the magnetic flux tubes are
also moving outward with them. According to this argument, 
Cheng \& Dai (1998) proposed a plate tectonic model for strange stars
which is, in principle, similar to that proposed by Ruderman (1991) 
for neutron stars. In this model, when the star spins down due to 
magnetic dipole radiation and wind emission, the vortex lines 
move outward and pull the flux tubes with them. However, 
since the terminations of the flux tubes are anchored in the base 
of the highly conducting crystalline crust, the flux tubes 
will produce sufficient tension to crack the crust and pull 
parts of the broken platelet into the strange quark matter. 
The time interval between two successive cracking events is
estimated to be (Cheng \& Dai 1998)
\begin{eqnarray}
\tau_{\rm int} & \sim & 10^6
\left(\frac{B_c}{10^{17}{\rm G}}\right)^{-1}
\left(\frac{B_p}{3\times 10^{13}{\rm G}}\right)^{-1}
\left(\frac{\theta_s}{0.03}\right) \nonumber \\
& & \times \left(\frac{\mu}{10^{27}{\rm dyn}\,{\rm cm}^{-2}}\right)
\left(\frac{l}{10^4{\rm cm}}\right)
\left(\frac{R}{10^6{\rm cm}}\right)^{-1}
\left(\frac{t_{\rm SNR}}{10^4{\rm yr}}\right)\,\,{\rm s},
\end{eqnarray}
where $\theta_s$ and $\mu$ are the shear angle (estimated below)
and the shear modulus ($\sim 10^{27}$ ${\rm dyn}\,{\rm cm}^{-2}$) 
at the base of the crust of the strange star respectively, and $l$ 
is the crustal thickness ($\sim 10^4$ cm). The melting temperature
of the crust is $T_m\approx 10^3(\rho_b/{\rm g}\,{\rm cm}^{-3})^{1/3}
Z^{5/3}\,{\rm K} \sim 10^9\,{\rm K}$, where 
$\rho_b$ is the mass density at the base of the crust 
($\sim 4\times 10^{11}\,{\rm g}\,{\rm cm}^{-3}$) and $Z$ is 
the charge number of nuclei ($Z=26$ for iron) (Shapiro \&
Teukolsky 1983). At age of $\sim 10^4$ yr, the interior temperature 
of the strange star $T\sim 10^7\,{\rm K}\ll 0.1T_m$ and thus
$\theta_s\sim 10^{-1}-10^{-2}$ (Ruderman 1991). We see that 
the time given by equation (1) is consistent with the typical time
interval between SGR bursts.

Because each baryon can release the deconfinement energy of 
$\sim 30$ MeV (the accurate value is dependent upon the quantum
chromodynamics parameters), the total amount of energy release
is estimated as
\begin{equation}
\Delta E_b\sim 3\times 10^{42} \left(\frac{\eta}{0.1}\right)
\left(\frac{M_{\rm cr}}{10^{-5}M_\odot}\right)
\left(\frac{l}{10^4{\rm cm}}\right)^2
\left(\frac{R}{10^6{\rm cm}}\right)^{-2} \,{\rm ergs},
\end{equation}
where $\eta$ is the fractional mass in the cracking area 
$\sim l^2$ which is dragged into the core (Cheng \& Dai 1998). 
At least half of this amount will be carried away by thermal
photons with the typical energy $kT\sim 30$ MeV. In the presence of 
a strong magnetic field ($\sim 3\times 10^{13}$ G), these thermal 
photons will convert into electron/positron pairs when
$[E_\gamma/(2m_ec^2)]B\sin \Theta/B_q\sim 1/15$,
where $E_\gamma$ is the photon energy, $B_q=m_e^2c^3/(\hbar e)
=4.4\times 10^{13}$ G, and $\Theta$ is the angle between the photon
propagation direction and the direction of the magnetic field
(Ruderman \& Sutherland 1975). The energies of the resulting pairs
will be lost via synchrotron radiation. The characteristic
synchrotron energy is given by
$E_{\rm syn}\sim 1.5\gamma_e^2\hbar eB\sin \Theta/(m_ec)
\sim 3.0\,\,{\rm MeV}$,
where $\gamma_e$ is the Lorentz factor of the pairs ($\sim 30$). 
These synchrotron photons will be converted into secondary pairs
because the optical depth for photon-photon pair production is
much larger than one. The Lorentz factor of the secondary pairs
is about 3.0. Thus, we obtain a cooling distribution of mildly
relativistic pairs, whose self-absorbed synchrotron emission 
has been shown to provide excellent fits to the spectral data 
of SGR bursts (Liang \& Fenimore 1995). In addition, after the
cracking event, roughly half of the resulting thermal energy
from deconfinement of normal matter to strange quark matter
will be absorbed by the stellar core, and thus the surface radiation 
luminosity at thermal equilibrium has been found to be consistent 
with the observed persistent X-ray luminosity (Cheng \& Dai 1998).    

As the quantized vortex lines move outward during the 
stellar spinning down and pull the magnetic flux tubes together, 
the magnetic toruses are also pulled toward the equatorial region
with the flux tubes due to the interaction between them
(Chau, Cheng \& Ding 1992). The upper limit of
$B_\phi$ is $B_f$; its lower limit can be given as follows. 
The magnetic toruses must are sufficiently buoyant 
to overcome fully the stratification in the composition 
of the crust, requiring $B_\phi \ge (8\pi \rho_bc_s^2)^{1/2}
\sim 5\times 10^{15}$ G, where $c_s$ is the speed of 
sound of the crust ($c_s\sim 2\times 10^9\,{\rm cm}\,
{\rm s}^{-1}$). The density of the flux tubes is given by 
$n=B_p/\Phi_v\sim 4\times 10^{20}\,{\rm cm}^{-2}$, 
where $\Phi_v$ is the magnetic flux of each tube ($\sim 8\times 
10^{-8}\,{\rm G}\,{\rm cm}^{-2}$) (Chau 1997). Now we define a 
timescale: $\Delta t=1/(v\sqrt{n})\sim 2\times 10^{-5}\,{\rm s}$,
where $v$ is the speed of flux tubes ($\sim 3\times 10^{-6}\,
{\rm cm}\,{\rm s}^{-1}$) (Cheng \& Dai 1998). The physical meaning
for $\Delta t$ is that there must be a flux tube to move toward
to the surface in the equatorial region in this timescale, implying
that there must be magnetic toruses which are pulled simultaneously to
the surface in the equatorial region and which break the surface.
The reconnection of magnetic toruses leads to 
an episodic wind. Because $\Delta t$ is much smaller than 
$\tau_{\rm int}$, too many episodic winds will constitute 
a steady wind. The magnetic energy including in each episodic 
wind can be estimated as $e_w= V_bB_\phi^2/(8\pi)$
where $V_b$ is the volume of a torus. It is important to note
that the magnetic torus is emerging from the quark matter core
to the surface and subsequently releasing this energy via 
reconnection in the equatorial region. 
Such an energy must be contaminated by crustal 
baryons whose mass is at most $m_w\sim M_{\rm cr}V_b/
(4\pi R^2l)$. In fact, the mass of the contaminating baryons
should be a fraction ($\xi$) of $m_w$. Thus, the Lorentz factor 
of the steady wind is
\begin{equation}
\Gamma\sim \frac{e_w}{\xi m_wc^2}
\sim 30 
\left(\frac{\xi}{0.1}\right)^{-1}
\left(\frac{M_{\rm cr}}{10^{-5}M_\odot}\right)^{-1}
\left(\frac{B_\phi}{10^{17}{\rm G}}\right)^2
\left(\frac{R}{10^6{\rm cm}}\right)^2
\left(\frac{l}{10^4{\rm cm}}\right).
\end{equation}
The luminosity of the wind can be estimated as 
\begin{equation}
L_w\sim \frac{B_\phi^2}{8\pi}\frac{\chi R^3}{t_{\rm SNR}}\sim 10^{37}
\left(\frac{\chi}{10^{-2}}\right)
\left(\frac{B_\phi}{10^{17}{\rm G}}\right)^2
\left(\frac{t_{\rm SNR}}{10^4{\rm yr}}\right)^{-1}{\rm erg}\,{\rm s}^{-1},
\end{equation}
where $\chi R^3$ is the volume of all the flux tubes ($\chi R$ is 
the characteristic length of the velocity gradient and is assumed to be
$\sim l$). This relativistic wind will inject into 
and power the surrounding SNR.
   
\begin{center}
4. DISCUSSION
\end{center}

Quiescence and burst emission and relativistic particle
winds of SGRs have been widely interpreted to result from
ultrastrongly magnetized neutron stars (Thompson \&
Duncan 1995, 1996). In such a magnetar model, the SGR bursts are 
due to readjustment of the magnetic field, possibly accompanied by
cracking of the neutron-star crust, and the persistent X-ray emission
is due to decay of the magnetic field, while the winds are due to thermal 
radiation from hot spots and Alfven wave emission (Thompson \& Blaes
1998). It is very clear that the energy sources of all the emission and
winds are the magnetic energy and gravitational energy of 
the neutron star. However, Harding et al. (1999) have shown that the
magnetic field must be up to $\sim 3\times 10^{13}$ G in order to
match the characteristic spin-down timescale for SGR $1806-20$
and its surrounding SNR age. Here we have argued that if the magnetic 
field is indeed so low, the previously suggested energy sources seem
problematic because they are much smaller than the observed energies 
for the persistent X-ray and burst emission and the wind. We have further 
proposed another plausible model in which SGR pulsars are young 
strange stars with superconducting cores and with magnetic
fields of $\sim 3\times 10^{13}$ G following Cheng \& Dai (1998)
and Dai \& Lu (1998). In our model, the movement of the flux tubes 
not only leads to crustal cracking, giving rise to deconfinement of
crustal matter to strange matter, but also to movement of internal
magnetic toruses with the flux tubes. As have shown in the above section,
the former process will result in burst and quiescence emission and the
latter process will produce relativistic winds which will power the
surrounding SNR. 
 
Another group of sources having periods and period derivatives similar
to SGRs are the AXPs, pulsating X-ray sources with periods in the range
$6 - 12$ s and period derivatives in the range of $10^{-12} - 10^{-11}
\,{\rm s}\,{\rm s}^{-1}$ (Gotthelf \& Vasisht 1998). These sources 
have shown only strong quiescent X-ray emission with no bursting 
behavior. Moreover, if the souces are highly magnetic pulsars, 
their characteristic ages ($P/2\dot{P}$) are in excellent agreement 
with the SNR ages, implying that the sources have no wind emission.
Why are there such obvious differences between SGRs and AXPs?
We suggest that AXPs be neutron stars with magnetic fields of 
$\sim 10^{15}$ G but without any toroidal magnetic field.
In the case of slowly rotating neutron stars, crustal cracking
cannot produce an observed burst because of too low 
gravitational energy release, and there is no wind 
in the absence of toroidal magnetic fields. 
  
This work was supported by a RGC grant of Hong Kong government 
and the National Natural Science Foundation of China.
   
\newpage
\baselineskip=4.0mm

\begin{center}
REFERENCES
\end{center}

\begin{description}
\item Bailin, B., \& Love, A. 1984, Phys. Rep., 107, 325
\item Baym, G., \& Pines, D. 1971, Ann. Phys., 66, 816
\item Benvenuto, O. G., Vucetich, H., \& Horvath, J. E. 1991,
        Nucl. Phys. B, 24, 160
\item Chau, H. F. 1997, ApJ, 479, 886
\item Chau, H. F., Cheng, K. S., \& Ding, K. Y. 1992, ApJ, 399, 213 
\item Cheng, K. S., \& Dai, Z. G. 1996, Phys. Rev. Lett., 77, 1210
\item Cheng, K. S., \& Dai, Z. G. 1998, Phys. Rev. Lett., 80, 18
\item Cheng, K. S., Dai, Z. G., \& Lu, T. 1998, Int. J. Mod. Phys. D, 7,
        139
\item Cheng, K. S., Dai, Z. G., Wei, D. M., \& Lu, T. 1998, Science, 
        280, 407
\item Cline, T. L. et al. 1982, ApJ, 255, L45
\item Dai, Z. G., \& Lu, T. 1998, Phys. Rev. Lett., 81, 4301
\item Dai, Z. G., \& Lu, T. 1999, ApJ, 519, L155
\item Duncan, R. C., \& Thompson, C. 1992, ApJ, 392, L2 
\item Evans, W. D. et al. 1980, ApJ, 237, L7
\item Fenimore, E. E., Laros, J. G., \& Ulmer, A. 1994, ApJ, 432, 742
\item Gotthelf, E. V., \& Vasisht, G. 1998, New Astron., 3, 293
\item Harding, A. K., Contopoulos, I., \& Kazanas, D. 1999, ApJ, 525, L125
\item Heyl, J. S., \& Kulkarni, S. R. 1998, ApJ, 506, L61
\item Hurley, K. et al. 1999, ApJ, 519, L143
\item Laros, J. G. et al. 1987, ApJ, 320, L111
\item Liang, E., \& Fenimore, E. E. 1995, ApJ, 451, L57
\item Kluzniak, W., \& Ruderman, M. A. 1998, ApJ, 505, L113
\item Kouveliotou, C. et al. 1994, Nature, 368, 125
\item Kouveliotou, C. et al. 1998, Nature, 393, 235
\item Kouveliotou, C. et al. 1999, ApJ, 510, L115
\item Kulkarni, S. R., \& Frail, D. A. 1993, Nature, 365, 33
\item Kulkarni, S. R. et al. 1994, Nature, 368, 129
\item Murakami, T. et al. 1994, Nature, 368, 127
\item Pacini, F. 1969, Nature, 221, 624
\item Paczynski, B. 1992, Acta Phys., 41, 145
\item Rothschild, R. E., Kulkarni, S. R., \& Lingenfelter, R. E. 1994,
        Nature, 368, 432
\item Ruderman, M. A. 1991, ApJ, 366, 261
\item Ruderman, M. A., \& Sutherland, P. G. 1975, ApJ, 196, 51
\item Shapiro, S. L., \& Teukolsky, S. A. 1983, Black Holes,
        White Dwarfs and Neutron Stars (John Wiley \& Sons), P. 90
\item Stergioulas, N., Kluzniak, W., \& Bulik, T. 1999, A\&A, in press
        (astro-ph/9909152)
\item Thompson, C., \& Blaes, O. 1998, Phys. Rev. D, 57, 3219
\item Thompson, C., \& Duncan, R. C. 1995, MNRAS, 275, 255
\item Thompson, C., \& Duncan, R. C. 1996, ApJ, 473, 322
\item Usov, V. V. 1992, Nature, 357, 452
\item Vasisht, G., Kulkarni, S. R., Frail, D. A., \& Greiner, J. 1994,
        ApJ, 431, L35 
\item Wang, X. Y., Dai, Z. G., Lu, T., Wei, D. M., \& Huang, Y. F. 1999,
        astro-ph/9910029
\item Woods, P. M. et al. 1999, ApJ, 519, L139
\item Zhang, W. et a. 1998, ApJ, 500, L171
\end{description} 

\end{document}